\documentstyle[twoside,12pt,epsf,amssymb]{article}
\begin{document}
\title{\bf Generalization of   Sato equation and systems of 
 multidimensional nonlinear Partial Differential Equations}

\author{ Alexandre I.
Zenchuk
\\Center of Nonlinear Studies of\\ L.D.Landau Institute 
for Theoretical Physics  \\
(International Institute of Nonlinear Science)\\
Kosygina 2, Moscow, Russia 117334\\
E-mail: zenchuk@itp.ac.ru
\\}
\maketitle

\begin{abstract}
This paper develops one  
of the methods for study of nonlinear  Partial Differential 
equations. We generalize Sato equation and represent the algorithm 
for construction of some classes of nonlinear Partial 
Differential Equations (PDE) together with solutions
parameterized by the set of arbitrary functions.
\end{abstract}

\section{Introduction}
Nonlinear Partial Differential Equations (PDE) have a great variety of application in both mathematical physics and applied mathematics. 
The well developed tools for study are admitted by the
 wide class of PDE which is called Completely Integrable Nonlinear PDE. These systems reflect many important properties of real physical objects and have been studied in many details. 
 So, Korteweg-de Vries equation (KdV)
\cite{GGKM,AKNS,ZMNP,WH}, Kadomtsev-Petviashvili equation (KP) \cite{WH}, 
Camassa-Holm equation \cite{CH1,CH2,FO,FF} 
and some of its two dimensional generalizations
(for instance, \cite{CGP}), are well known in hydrodynamics, where 
they describe propagation of  surface waves on incompressible fluid 
without viscosity; Nonlinear 
Schr\"odinger equation (NLS) is closely used in  nonlinear fiber optics, where 
 it describes propagation of electromagnetic pulses \cite{HKO,A}. 
Note that some  methods for study of the integrable PDE 
do not presume in advance their integrability.
For instance,
Sato theory \cite{OSTT,M,P}, Hirota bilinear method \cite{H1,H2,HS,HS2}, Penlev\'e method \cite{WTC,W,EG}, recent modification of the dressing method 
based on the algebraic matrix equation \cite{Z1,Z2,Z3}.

In this paper 
we use the ideas of the ref.\cite{OSTT}, where the space of
one dimensional pseudo-differential (or differential)
operators has been used for description of
the KP hierarchy. We replace the space of 
one dimensional operators with the space of 
 multi-dimensional {\it differential} operators, which allows one to
 treat larger class of  PDE, in particular, ($n+1$)-dimensional PDE 
with $n>2$.  
We  provide the  family of particular solutions, 
parameterized by the set of arbitrary functions.
 Although we have not answered the question about complete integrability of the 
PDE under consideration. 
In comparison  with approach considered in the refs. \cite{Z1,Z2,Z3}, 
the manifold of available solutions is  reacher. 
Trying to make the understanding of this article independent on references, 
we supply all necessary details below. 

General idea is developed in Sec.2. The useful subspace of $n$-order
multidimensional differential operators together with related systems of 
nonlinear PDE is considered in the Sec.3. We represent simple example of 
(2+1)-dimensional equation and generalization of the KP hierarchy in the same section.
Algorithm for construction of the families of particular solutions is discussed in the Sec.4. Conclusions are given in the Sec.5.

\section{Space of $N$-dimensional 
differential operators and related systems of nonlinear PDE}

Hereafter we will use 
 Greek letters to denote multi-index, 
$\alpha=(\alpha_1\dots\alpha_N)$, where $N$ is the
dimension of $x$-space. The dimension of $t$-space (space of additional
parameters $t_\gamma$, $\gamma=(\gamma_1,\dots \gamma_N)$ )
may be arbitrary. In particular, it may be infinity.
By definition,
\begin{eqnarray}\label{def}
x=(x_1,\dots,x_N),\;\;
||\alpha||=\sum_{i=1}^N\alpha_i,\;\;\\\nonumber
\partial^\alpha=\prod_{i=1}^N
\partial_i^{\alpha_i},\;\;
\partial_i\equiv\frac{\partial}{\partial {x_i}},\;\;
\partial_\gamma\equiv \frac{\partial}{\partial t_\gamma},
\end{eqnarray}
\begin{eqnarray}\label{Qaln}
Q^{\alpha}_n \;\; {\mbox{is the number of all possible values of 
 }}\;\;
\alpha \;\;{\mbox{with }}\;\;||\alpha||=n.
\end{eqnarray}
All functions are differentiable with respect to their arguments
as many times as one needs.

Denote ${\Bbb P}^n$ 
the linear  space of $n$-order differential operators of the following general 
form
\begin{eqnarray}\label{W}
W_n \in {\Bbb P}^n : W_n = 
\sum_{||\beta||=0}^{n} w_\beta \partial ^\beta,
\end{eqnarray}
where $w_\beta$ are  functions of $x$ and $t$, which will
be specified below. Denote $Q_n$ the  number of terms
 in the operator $W_n$ of general form (\ref{W}):
 \begin{eqnarray}\label{Qn}
 Q_n=\sum_{||\alpha||=0}^n Q^\alpha_n.
 \end{eqnarray}
Space ${\Bbb P}^n$ is related with  $Q_n$-dimensional 
linear space ${\Bbb C}^{Q_n}$ of
 functions $\psi(x,t)$ with the basis
$\psi_i$, $i=1,\dots, Q_n$, satisfying the following condition:
\begin{eqnarray}\label{det}
{\mbox{det}}||\partial^\beta \psi_i|| \neq 0 ,\;\;
0\le||\beta||\le n,\;\;i=1,\dots,Q_n,
\end{eqnarray}  
where $||\partial^\beta \psi_i||$ is $Q_n\times Q_n$ matrix. Indexes
$\alpha $ and $i$ enumerate columns and rows of this matrix respectively.
By definition, dependence on parameters $x$ is arbitrary, while
parameters $t_\gamma$
are introduced in functions $\psi_i$  by the following formulas:
\begin{eqnarray}\label{t}
\partial_\gamma \psi_i =
\partial^\gamma \psi_i.
\end{eqnarray}

{\bf Suggestion 1.} Let parameter  $Q_n$ 
be defined by the eq.(\ref{Qn}). For any $\alpha$ such that
$||\alpha||=n$ and
any space 
${\Bbb C}^{Q_{n-1}}$
 there is
set of $Q_{n-1}$  {\it uniquely} 
defined 
functions $w^\alpha_\beta$ such that 
space ${\Bbb C}^{Q_{n-1}}$ forms $Q_{n-1}$-dimensional linear subspace of
solutions for the following  linear differential equation:
\begin{eqnarray}\label{Seq}
S_\alpha \psi =0, \;\;S_\alpha =  \partial^\alpha + W_\alpha,\;\;
W_\alpha=\sum_{||\beta||=0}^{n-1} w^\alpha_\beta 
 \partial^\beta \in {\Bbb P}^n.
\end{eqnarray}

$\vartriangle$
Let us fix parameter $\alpha$ with $||\alpha||=n$ and
 basis $\psi_i$, $i=1,\dots,Q_{n-1}$ of space
${\Bbb C}^{Q_{n-1}}$. One can write  the 
formal system of differential equations
(\ref{Seq})
\begin{eqnarray}
\label{Su1} 
S_\alpha\psi_i\equiv \partial^\alpha\psi_i + 
\sum_{||\beta||=0}^{n-1} w^\alpha_\beta
\psi_i=0,\;\;i=1,\dots,Q_{n-1}.
\end{eqnarray} 
Let us consider the  system (\ref{Su1}) as nonhomogeneous 
 system of {\it algebraic} equations on the functions $w^\alpha_\beta$. The
 matrix of this system,
 $||\partial^\beta \psi_i||$ ($0\le ||\beta||\le
 n-1$, $i=1,\dots,Q_{n-1}$), is nondegenerate due to the condition
 (\ref{det}). So that this system has {\it unique} solution,
 i.e. functions $w^\alpha_\beta$ are defined uniquely in terms of
 functions $\psi_i$ and their derivatives. Owing to this,
 $w^\alpha_\beta$ are  functions of both sets of variables, $x$
 and $t$. By construction, we have determined coefficients
 $w^\alpha_\beta$ in the system of differential equations
 (\ref{Seq}), for which given set of functions $\psi_i$
 ($i=1,\dots,Q_{n-1}$) forms basis for certain solution subspace.
$\blacktriangle$

Equations of the form (\ref{Seq}) with all possible values of
parameter 
$\alpha$ such that $||\alpha||=n$, form system of  $Q^\alpha_n$ differential equations. 
One can show that its coefficients   $w^\alpha_\beta$ 
satisfy some system of nonlinear PDE. 
For this purpose, one needs to fix multi-index $\gamma$,
$||\gamma||=n_\gamma$, 
differentiate the eqs.(\ref{Seq}) with respect to $t_\gamma$
and study the result:
\begin{eqnarray}\label{St}
\left(\frac{\partial S_\alpha}{\partial t_\gamma} + 
S_\alpha \partial^\gamma\right)\psi=0,\;\;||\gamma||=n_\gamma. 
\end{eqnarray}

{\bf Suggestion 2.} For each particular pair of parameters
$\alpha$ ($||\alpha||=n$) and $\gamma$ 
there is set of $Q^\alpha_n$  differential
operators 
$B_{\gamma\alpha\epsilon}=\sum_{||\delta||=0}^{n_\gamma}
v_{\gamma\alpha\epsilon\delta}\partial^\delta \in 
{\Bbb P}^{n_\gamma}$ 
with coefficients expressed in
terms of functions $w^\alpha_\beta$ and their derivatives
such that operators $M_{\gamma\alpha}$
\begin{eqnarray}
\label{M}
M_{\gamma\alpha} \equiv \frac{\partial S_\alpha}{\partial t_\gamma} + 
S_\alpha \partial^\gamma +
\sum_{||\epsilon||=n} B_{\gamma\alpha\epsilon} S_\epsilon  
\end{eqnarray}
belong to the space ${\Bbb P}^{n-1}$,
i.e. $M_{\gamma\alpha}$ can be written in the form 
\begin{eqnarray}\label{Meq}
M_{\gamma\alpha} = 
\sum_{||\beta||=0}^{n-1} a_{\beta}^{\gamma\alpha} \partial^\beta,
\end{eqnarray} 
where $a_{\beta}^{\gamma\alpha}$ are some functions of
$\omega^\alpha_\beta$ and their derivatives.

$\vartriangle$
The first term in the operator $M_{\gamma\alpha}$,
$\displaystyle \frac{\partial S_\alpha}{\partial t_\gamma}\equiv
\sum_{||\beta||=0}^{n-1} \frac{\partial w^\alpha_\beta}{\partial
t_\gamma}\partial^\beta$,
belongs to the space ${\Bbb P}^{n-1}$.
Since $S^\alpha\in {\Bbb P}^n$, the second term in the eq.(\ref{M})
belongs to the space 
${\Bbb P}^{n+n_\gamma}$.
Let $ B_{\gamma\alpha\epsilon}=\sum_{||\delta||=0}^{n_\gamma}
 v_{\gamma\alpha\epsilon\delta}\partial^\delta \in 
{\Bbb P}^{n_\gamma}$ be  $n_\gamma$-order 
differential operators with arbitrary coefficients. 
Then  one can write
\begin{eqnarray}
 \sum_{||\epsilon||=n}
 B_{\gamma\alpha\epsilon}
S_\epsilon &=& \sum_{||\beta||=n}^{n+n_\gamma} 
b_{\gamma\alpha\beta} \partial^\beta + R^{n-1},\;\;
R^{n-1}\in {\Bbb P}^{n-1},\\
b_{\gamma\alpha\beta}&=&
v_{\gamma\alpha\epsilon(\beta-\epsilon)} +\\\nonumber
&& f(
v_{\gamma\alpha\epsilon\tilde\delta},\;
||\tilde\delta||>||\beta-\epsilon||;
{\mbox{coefficients of operators $W_\alpha$}}),\;\;\\\nonumber
&& ||\beta||=n,\dots,n+n_\gamma.
\end{eqnarray}
 This means that $\sum_{||\epsilon||=n}
B_{\gamma\alpha\epsilon}
S_\epsilon\in {\Bbb P}^{n+n_\gamma}$ and  has all arbitrary 
coefficients ahead of the
derivatives of the order grater or equal to $n$. 
Let us  choose these coefficients in such a way that 
\begin{eqnarray}\label{SP}
\left(S_\alpha \partial^\gamma +
\sum_{||\beta||=n}  B_{\gamma\alpha\epsilon} S_\epsilon\right) 
\in {\Bbb P}^{n-1}.
\end{eqnarray}
 which 
provides condition
 $M_{\gamma\alpha}\in {\Bbb P}^{n-1}$.
 Due to the relation (\ref{SP}), coefficients 
 $v_{\gamma\alpha\epsilon\delta}$ are expressed in terms of the
 coefficients of the operators $W_\alpha$, $||\alpha||=n$. 
$\blacktriangle$

By construction, operators $M_{\gamma\alpha}$ generate the following 
system of $Q_{n-1} \times Q^\alpha_n$ linear differential equations:
\begin{eqnarray}\label{MSyst}
M_{\gamma\alpha} \psi_i \equiv \sum_{||\beta||=0}^{n-1} 
a^{\gamma\alpha}_\beta \partial^\beta \psi_i= 0,\;\;||\alpha||=n
,\;\;i=1,\dots,Q_{n-1}.
\end{eqnarray}
Let us regard the system (\ref{MSyst}) as linear {\it homogeneous}
system of algebraic  equations
on coefficients $a^{\gamma\alpha}_\beta$ with nondegenerate
matrix $||\partial^\beta\psi_i||$ ($0\le ||\beta||\le n-1$) due
to the condition (\ref{det}). 
This system  has only trivial
solution, i.e.
\begin{eqnarray}\label{Su2}
a^{ \gamma\alpha}_{\beta} &=& 0.
\end{eqnarray}
From another point of view, coefficients 
$a^{\gamma\alpha}_{\beta}$ have been defined in terms of
coefficients of the operators $W_\alpha$ and their 
derivatives, i.e.
the system  (\ref{Su2}) is system of nonlinear PDE on functions
$w^\alpha_\beta$, generated by  the space of  $(n-1)$-order
differential operators $W_{\alpha}$  with parameter $t_\gamma$ introduced by
the eq.(\ref{t}). 
The complete  set of systems of nonlinear PDE, related with 
 all possible parameters $t_\gamma$, introduced by
the eq.(\ref{t}), forms hierarchy of PDE associated with given space 
${\Bbb P}^{n-1}$.

We call the equation (\ref{MSyst}) written in the form
\begin{eqnarray}
\label{SB}
\frac{\partial S_\alpha}{\partial t_\gamma} + 
S_\alpha \partial^\gamma +
\sum_{||\epsilon||=n} B_{\gamma\alpha\epsilon} S_\epsilon =
0,\;\;||\alpha||=n
\end{eqnarray}
 {\it generalization of Sato equation}.
The system of equations (\ref{Su1}) serves 
for construction of particular solutions to the system
of nonlinear PDE
(\ref{Su2}),
which will be used in Sec.4.

 The system (\ref{Su2})
consists of  $Q_{n-1}\times  Q^\alpha_n$ equations.
Equations of the system (\ref{Su2})
with any particular $\gamma$  
recursively relate all $Q_{n-1}\times  Q^\alpha_n$ coefficients 
$w^\alpha_\beta$ ($||\alpha||=n$, $||\beta||=0,1,\dots,n-1$) 
of the operators $W_\alpha$ and compose the complete system of 
equations for whole set of these coefficients. 
Disadvantage of this system is that  its structure  depends on
$n$.
 In the next section 
we consider the subspace of the space ${\Bbb P}^{n-1}$, 
which allows one to generate the systems of PDE  
 which do not depend on the value of parameter 
$n$, if only this value is big enough. These systems of PDE
admit the 
class of solutions, which depends on the  set of 
 arbitrary functions of single variable.

\section{About reductions}

Let ${\cal P}^n_{n_2,\dots,n_N} \subset {\Bbb P}^n$, 
$n=\sum_{i=1}^N n_i$, be defined as follows:
\begin{eqnarray}\label{Wred}
W_n \in {\cal{ P}}^n_{n_2,\dots,n_N} : W_n = 
\sum_{||\beta||=0}^{n} w_{\hat \beta} \partial ^\beta, \;\;
\beta_i\le n_i \;\;{\mbox{for}}\;\; i>1,\;\;\\\nonumber 
\beta=(\beta_1\beta_2 \dots\beta_N),\;\;
\hat \beta=((n-\beta_1-1)\beta_2 \dots\beta_N),
\end{eqnarray}
i.e. parameters $n_i$, $i=1,\dots,N$, mean the order of 
the highest derivative with respect to $x_i$
in the differential operators $W_n$. Emphasize that we don't put any restriction on the 
order of the derivative with respect to $x_1$, i.e its highest possible 
order in operators $W_n$  equals $n$. 
Index with hat is introduced for convenience of representation of
system of nonlinear PDE.
All multi-indexes (without hat) in this section have 
the following structure:
\begin{eqnarray}\label{hat}
\alpha=(\alpha_1\dots\alpha_N)\;\;{\mbox{ with}}\;\; 
\alpha_i<n_i\;\;{\mbox{ for}}\;\;i>1.
\end{eqnarray} 

Let us 
introduce the set of equations, which specifies the dependence on $x_i$, $i>1$:
\begin{eqnarray}\label{x}
\partial_i^{n_i+1}\psi = \partial^{\alpha^{(i)}}\psi,\;\;
\alpha^{(i)}=(\alpha^{(i)}_1\dots \alpha^{(i)}_N),\;\;
\alpha^{(i)}_i=0,
\end{eqnarray}
where $\psi\in{\Bbb C}^{Q_n}$. Recall that $Q_n$ is the maximum  
number of terms in 
operators $W_n\in {\cal{ P}}^n_{n_2,\dots,n_N}$.  
Due to the relations (\ref{x}), one can use variables $x_i$ ($i>1$) along with $t_\gamma$  
for construction of the nonlinear eqs. (\ref{Su2}).
 One needs just replace $\partial_\gamma$ with 
$\partial_i$ in formulas (\ref{St}-\ref{SB}). 
This fact will be used in sections 3.1.1 and 3.1.2.

\subsection{Examples of nonlinear PDE related with space $
{\cal{ P}}^n_1$}
 
In this section we will use two-dimensional $x$-space, $x=(x_1,x_2)$, 
$W_j\in {\cal P}^{n-1}_1$, $j=1,2,\dots$. 
Maximum order of derivative with respect
to $x_2$  in the operators 
$W_j$ equals one: 
$n_2=1$. 
Because of that,
the set of operators $S_\alpha$ consists of two operators
 $S_1\equiv S_{n0}$ and $S_2\equiv S_{(n-1)1}$ for any fixed parameter $n$:
\begin{eqnarray}\label{S1}
S_1 \equiv \partial_1^n +W_1,\;\;
W_1=\sum_{k=0}^{n-2} u_{(n-k-2) 1} \partial_{1}^k\partial_{2}+
     \sum_{k=0}^{n-1} u_{(n-k-1) 0}\partial_{1}^k,  \\\label{S2}
S_2 \equiv \partial_1^{n-1}\partial_2 +W_2,\;\; W_2= \sum_{k=0}^{n-2} 
   v_{(n-k-2) 1} \partial_{1}^k\partial_2+
\sum_{k=0}^{n-1} v_{(n-k-1) 0}\partial_{1}^k.  
\end{eqnarray}
These operators introduce two differential equations (\ref{Seq})
on the function $\psi$:
\begin{eqnarray}\label{Seq2}
S_1 \psi = 0,\;\;\;S_2\psi=0.
\end{eqnarray}
It is simple to define the number of terms $Q_{n-1}$  in the
 operators $W_{j}$:
 \begin{eqnarray}
 \label{Q}
 Q_{n-1}=2n-1. 
\end{eqnarray}
Next, one needs to fix $(2n-1)$-dimensional basis 
$\psi_i$ ($i=1,\dots, 2 n - 1$) in subspace ${\Bbb C}^{2 n-1}$
and write formally the system (\ref{Su1})
\begin{eqnarray}\label{Seq2i}
S_1\psi_i&=&0,\;\;\\\label{Seq2ii}
S_2\psi_i&=&0,\;\;i=1,\dots,2n-1,
\end{eqnarray} 
Condition (\ref{det}) now reads
{\small
\begin{eqnarray}\label{det2}
 {\small \left | 
\begin{array}{ccccccc}
\partial_1^{n-1} \psi_1&\partial_1^{n-2} \psi_1&\cdots&
\psi_1&\partial_1^{n-2} \partial_2\psi_1&
\cdots& \partial_2\psi_1\cr
\cdots&\cdots&\cdots&\cdots&\cdots&\cdots&\cdots\cr
\partial_1^{n-1} \psi_{2 n-1}&\partial_1^{n-2} \psi_{2 n-1}&\cdots&
\psi_{2 n-1}&\partial_1^{n-2} \partial_2\psi_{2 n-1}&
\cdots& \partial_2\psi_{2 n-1}\end{array}\right|}\neq 0
\end{eqnarray}
}
Each of the systems  (\ref{Seq2i})  and (\ref{Seq2ii}) 
should be treated as system of algebraic equations 
on  coefficients $u_{ij}$
and 
$v_{ij}$ respectively.
Due to the condition (\ref{det2}) these equations have unique
nontrivial solution. 

\subsubsection { Example 1: (2+1) dimensional PDE}

In this example 
we take the eq.(\ref{x}) in the following form 
\begin{eqnarray}\label{x2}
\partial_{2}^2\psi= \partial_1\psi,\;\;
\end{eqnarray}
and use notations 
\begin{eqnarray}\label{def3}
x\equiv x_1,\;\;t_1\equiv x_2,\;\;t_2\equiv t_{\alpha},\;\;\alpha=(20).
\end{eqnarray}
Function $\psi$ depends on   variable $t_2$ due to the formula 
(see eq. (\ref{t}))
\begin{eqnarray}
\partial_{t_2}\psi=\partial_1^2 \psi.
\end{eqnarray}
Equations (\ref{SB}) related with parameters $t_1\equiv x_2$ 
and $t_2$ 
have the following form:
\begin{eqnarray}
M_{1j}  = 
\frac{\partial S_j}{\partial x_2} +S_j\partial_1 + 
B_{1j1} S_1 + B_{1j2} S_2=0, \;\; j=1,2,\\
M_{2j}  = 
\frac{\partial S_j}{\partial t_2} +S_j\partial_1^2 + 
B_{2j1} S_1 + B_{2j2} S_2=0, \;\; j=1,2,
\end{eqnarray}
where $B_{ijk}$ are  differential operators 
 of the 
next form
\begin{eqnarray}\label{B1}
B_{1 1 1} &=& v_{0 0},\;\; B_{1 1 2}\; =\; 
   (-u_{0 0} + v_{0 1}) - \partial_1,\;\; \\
  B_{1 2 1} &=& -1,\;\; B_{1 2 2} \;=\; -v_{0 0},\;\; \\
  B_{2 1 1}& =& 
   2 { u_{0 0}}_x - \partial_1^2,\;\; 
  B_{2 1 2} \;=\; 2 { u_{01}}_x,\;\; \\\label{B2}
  B_{2 2 1} &=& 2 { v_{00}}_x,\;\; 
  B_{2 2 2} \;=\; 
   2 {v_{01}}_x - \partial_1^2,
\end{eqnarray}
which provide the following structure of the operators 
$M_{ij}$ (see (\ref{Meq})):
\begin{eqnarray}\label{MU2}
M_{ij}=\sum_{s=0}^{n-2} a^{ij}_{s1}\partial_1^s \partial_2 + 
\sum_{s=0}^{n-1} a^{ij}_{s0}\partial_1^s ,\;\;i,j=1,2
\end{eqnarray}
and
\begin{eqnarray}\label{nonlin}
a^{ij}_{sk} \equiv 0
\end{eqnarray}
(compare with derivation of the eq.(\ref{Su2})). 

Below are  several equations from the list (\ref{nonlin}) which
compose the complete system of equations:
\begin{eqnarray}\label{U1}
a^{1 1}_{ (n-1) 0} &=& 
   u_{0 1} + v_{0 0}v_{0 1} - v_{1 0} + 
   { u_{00}}_{x_2} -  {v_{00}}_x=0,\;\; \\\label{U2}
  a^{1 1}_{ (n-2) 1} & =&  
   u_{1 0} + u_{0 1}v_{0 0} - u_{0 0}v_{0 1} + v_{0 1}^2 - v_{1 1} + 
     {u_{01}}_{x_2} - {v_{01}}_x=0,\;\; \\\label{U3}
  a^{1 2}_{(n-1) 0} & = & 
   -u_{0 0} - v_{0 0}^2 + v_{0 1} +  {v_{00}}_{x_2}=0,\;\; \\\label{U4}
  a^{1 2}_{( n-2) 1} & =&  
   -u_{0 1} - v_{0 0}v_{0 1} + v_{1 0} + {v_{01}}_{x_2}=0,\;\; 
\\\label{U5}
  a^{1 2}_{ (n-2) 0} & =&  
   -u_{1 0} - v_{0 0}v_{1 0} + v_{1 1} + {v_{10}}_{x_2}=0,\;\; \\
\label{U6}
  a^{2 1}_{ (n-1) 0} & =&  
    {u_{00}}_{t_2} + 
    2u_{0 0}{u_{00}}_x + 
    2v_{0 0}{u_{01}}_x - 
    2 {u_{10}}_x -  {u_{00}}_{xx}=0
,\;\;\\\label{U7}
 a^{2 1}_{ (n-2) 1}&  =&  
   {u_{0 1}}_{t_2} + 
    2u_{0 1} {u_{00}}_x + 
    2v_{0 1} {u_{01}}_x - 
    2 {u_{11}}_x - {u_{01}}_{xx}=0
,\;\;\\\label{U8}
 a^{2 2}_{ (n-1) 0} & = & 
   {v_{00}}_{t_2} + 
    2u_{0 0} {v_{00}}_x + 
    2v_{0 0} {v_{01}}_x - 
    2 {v_{10}}_x -  {v_{00}}_{xx}=0
,\;\; \\\label{U9}
a^{2 2}_{ (n-2) 1}&  =&  
    {v_{0 1}}_{t_2} + 
    2u_{0 1} {v_{00}}_x + 
    2v_{0 1} {v_{01}}_x - 
    2 {v_{11}}_x - {v_{01}}_{xx}=0,
\end{eqnarray}
We leave without proof the statement that this system remains the
same for any value of parameter $n$, which happens due to the
special structure of the subspace ${\cal{P}}^{n-1}_{1}$.

To simplify the system (\ref{U1}-\ref{U9}), 
let us solve eqs. (\ref{U3},\ref{U4},\ref{U5})
with respect to $u_{00}$, $u_{01}$ and $u_{10}$ respectively, substitute the result  in the rest of equations. Then eqs.(\ref{U8}) and (\ref{U1},\ref{U2})
result in  the following system of equations on the functions  
$u=v_{00}$, $v=v_{01}$,
and $w=v_{10}$:
\begin{eqnarray}\label{uvw}
u_{t_2} - 
   2u^2 u_x + 
   2v u_x + 
   2 u_{x_2} u_x +
   2u v_x - 
   2 w_x - u_{xx}&=&0,\;\; \\\nonumber
  -2u u_{x_2} + 
   2 v_{x_2} + 
   u_{x_2x_2} -  u_x&=&0,\;\; \\\nonumber
\label{u4}
  -2v u_{x_2} + 
   2  w_{x_2} + 
   v_{x_2x_2} -  v_x&=&0.
\end{eqnarray}
Simple example of particular solution for this system will be constructed 
in the Sec.4.

\subsubsection{Example 2: (3+1)-dimensional generalization of KP hierarchy}

In this example we use the following notations:
\begin{eqnarray}\label{def4}
x\equiv x_1,\;\;t_1\equiv x_2,\;\; t_i\equiv t_{\alpha^{(i)}},\;\;i=2,3,\;\;
\alpha^{(2)}=(11),\;\;\alpha^{(3)}=(30).
\end{eqnarray}
Dependence on $t_1=x_2$, $t_2$ and $t_3$ is given by the following equations:
\begin{eqnarray}
\partial_2^2\psi=\partial_1^2\psi,\;\;
\partial_{t_2}\psi=\partial_1\partial_2\psi,\;\;
\partial_{t_3}\psi=\partial_1^3\psi,\;\;
\end{eqnarray}
Equations (\ref{SB}) have the following form:
\begin{eqnarray}\label{M21}
M_{1j}  = 
\frac{\partial S_j}{\partial x_2} +S_j\partial_1^{2} + 
B_{1j1} S_1 + B_{1j2} S_2=0, \;\; j=1,2,\\\label{M22}
M_{2j}  = 
\frac{\partial S_j}{\partial t_2} +S_j\partial_1 + 
B_{2j1} S_1 + B_{2j2} S_2=0, \;\; j=1,2,\\\label{M23}
M_{3j}  = 
\frac{\partial S_j}{\partial t_3} +S_j\partial_1 + 
B_{3j1} S_1 + B_{3j2} S_2=0, \;\; j=1,2,
\end{eqnarray}
One has the following expressions
for differential operators $B_{ijk}$:
\begin{eqnarray}
B_{1 1 1} &=&  (-u_{0 1} + v_{0 0}),\\ 
  B_{1 1 2}&=&  (-u_{0 0} + v_{0 1}) - \partial_1,\\ 
  B_{1 2 1}&=& (u_{0 0} - v_{0 1}) - \partial_1,\\ 
  B_{1 2 2} &=&  (u_{0 1} - v_{0 0}),\\ 
  B_{2 1 1} &=& 
        (u_{0 0} u_{0 1} - u_{1 1} - v_{0 0} v_{0 1} + v_{1 0} + 
       2 {v_{00}}_x)+(-u_{0 1} + v_{0 0}) \partial_1 ,\\ 
  B_{2 1 2} &=& (u_{0 1}^2 - u_{1 0} - u_{0 1} v_{0 0} + u_{0 0} v_{0 1} - 
       v_{0 1}^2 + v_{1 1} + 2 {v_{01}}_x) +\\\nonumber&& 
   (-u_{0 0} + v_{0 1}) \partial_1 -
   \partial_1^2,\\ 
  B_{2 2 1} &=& (-u_{0 0}^2 + u_{1 0} - u_{0 1} v_{0 0} + v_{0 0}^2 + 
       u_{0 0} v_{0 1} - v_{1 1} + 2 {u_{00}}_x)+\\\nonumber&&
   (u_{0 0} - v_{0 1}) \partial_1 -
   \partial_1^2,\\ 
  B_{2 2 2} &=& 
        (-(u_{0 0} u_{0 1}) + u_{1 1} + v_{0 0} v_{0 1} - v_{1 0} + 
       2 {u_{01}}_x)+(u_{0 1} - v_{0 0}) \partial_1,\\ 
  B_{3 1 1}& =& 
    - 
    3  (u_{0 0} {u_{00}}_x + 
       v_{0 0} {u_{01}}_x - 
       {u_{10}}_x - 
       {u_{00}}_{xx}) +3 {u_{00}}_x\partial_1 - \partial_1^3 ,\\
  B_{3 1 2} &= &
   - 
    3  (u_{0 1} {u_{00}}_x + 
       v_{0 1} {u_{01}}_x - 
       {u_{11}}_{x} - 
       {u_{01}}_{xx})+ 3  {u_{01}}_x \partial_1,\\ 
  B_{3 2 1} &=& 
    - 
    3  (u_{0 0} {v_{00}}_x + 
       v_{0 0} {v_{01}}_x - 
       {v_{10}}_{x} - 
       {v_{00}}_{xx})+3 {v_{00}}_x \partial_1,\\ 
  B_{3 2 2}& =& 
    - 
    3  (u_{0 1} {v_{00}}_x + 
       v_{0 1} {v_{01}}_x - 
       {v_{11}}_{x} - 
       {v_{01}}_{xx})+3  {v_{01}}_x\partial_1 -\partial_1^3.
\end{eqnarray}
With given operators $B_{ijk}$, 
 equations (\ref{M21}-\ref{M23}) take the form (\ref{MU2}) with 
$i=1,2,3$ and $j=1,2$.
Write down several nonlinear equations, generated by each of the
eqs.(\ref{M21}-\ref{M23}).

From  the eq.(\ref{M21}):
\begin{eqnarray}\label{u21}
a^{1 1}_{(n-1) 0} &=& 
   -(u_{0 0} u_{0 1}) + u_{1 1} + v_{0 0} v_{0 1} - v_{1 0} + 
    {u_{00}}_{x_2} - {v_{00}}_x=0,\\ \label{u22}
  a^{1 1}_{(n-2) 1} &=& 
   -u_{0 1}^2 + u_{1 0} + u_{0 1} v_{0 0} - u_{0 0} v_{0 1} + v_{0 1}^2 - 
    v_{1 1} +\\\nonumber && {u_{01}}_{x_2} - 
    {v_{01}}_x=0,\\\nonumber
  &&  ..........
\end{eqnarray}

From the eq.(\ref{M22}):
\begin{eqnarray}\label{u23}
  a^{2 1}_{ (n-1) 0}& =& 
   u_{0 0}^2 u_{0 1} - u_{0 1} u_{1 0} - u_{0 0} u_{1 1} + u_{2 1} + 
    u_{0 1}^2 v_{0 0} - u_{0 1} v_{0 0}^2 -\\\nonumber &&
 v_{0 0} v_{0 1}^2 + 
    v_{0 1} v_{1 0} + v_{0 0} v_{1 1} - v_{2 0} + 
    {u_{00}}_{t_2} - 
    u_{0 1} {u_{00}}_x + \\\nonumber &&
    v_{0 0} {u_{00}}_x + 
    u_{0 0} {v_{00}}_x +
    v_{0 1} {v_{00}}_x + 
    2 v_{0 0} {v_{01}}_x - \\\nonumber &&
    2 {v_{10}}_{x} - {v_{00}}_{xx}
=0,\\ 
a^{2 1}_{ (n-2) 1} &=& 
   u_{0 0} u_{0 1}^2 - 2 u_{0 1} u_{1 1} + u_{2 0} + u_{1 1} v_{0 0} + 
    u_{0 1}^2 v_{0 1} - \\\nonumber &&u_{1 0} v_{0 1} -
 2 u_{0 1} v_{0 0} v_{0 1} + 
    u_{0 0} v_{0 1}^2 - v_{0 1}^3 + u_{0 1} v_{1 0} -
\\\nonumber && u_{0 0} v_{1 1} + 
    2 v_{0 1} v_{1 1} - v_{2 1} +
{u_{01}}_{t_2} - 
    u_{0 1} {u_{01}}_x + 
    v_{0 0} {u_{01}}_x + \\\nonumber &&
    2 u_{0 1} {v_{00}}_x - 
    u_{0 0} {v_{01}}_x + 
    3 v_{0 1} {v_{01}}_x - 
    2 {v_{11}}_{x} - {v_{01}}_{xx}=0,\\\nonumber
 &&   ..........
\end{eqnarray}

From the eq.(\ref{M23}):
\begin{eqnarray} \label{u26}
  a^{3 1}_{ (n-1) 0}& =& 
   {u_{00}}_{t_3} - 
    3 u_{0 0}^2 {u_{00}}_x + 
    3 u_{1 0} {u_{00}}_x - 
    3 u_{0 1} v_{0 0} {u_{00}}_x + 
    3 {u_{00}}_x^2 - \\\nonumber&&
    3 u_{0 0} v_{0 0} {u_{01}}_x - 
    3 v_{0 0} v_{0 1} {u_{01}}_x + 
    3 v_{1 0} {u_{01}}_x + 
    3 u_{0 0} {u_{10}}_x + \\\nonumber&&
    3 v_{0 0} {u_{11}}_{x} - 
    3 {u_{20}}_{x} + 
    3 {u_{01}}_x 
     {v_{00}}_x + 
    3 u_{0 0} {u_{00}}_{xx} + \\\nonumber&&
    3 v_{0 0} {u_{01}}_{xx} - 
    3 {u_{10}}_{xx} - {u_{00}}_{xxx}
=0,\\ \nonumber
&&..........
\end{eqnarray}

Let us show that the above lists of nonlinear PDE can be reduced 
to the integrated KP. First of all note that 
the reduction $\partial_{2}\equiv \partial_1$ leads to the 
following identities:
\begin{eqnarray}
v_{ij}\equiv u_{ij},\;\; u_{i1}\equiv u_{i0},\;\;\mbox {for all} \;\;i
\;\;\mbox {and} \;\; j.
\end{eqnarray}
Then equations (\ref{u21}) and (\ref{u22}) becomes identities, 
while the system (\ref{u23}) - (\ref{u26}) results in 
 three equations (
$u=u_{00}$, $v=u_{10}$, $w=u_{20}$)
\begin{eqnarray}
u_{t_1} + 4 u u_x -2 v_x -u_{xx}=0,\;\;
v_{t_1} + 4 v u_x -2 w_x -v_{xx}=0,\\\label{kp}
u_{t_3} - u_{xxx} - 3 v_{xx}- 3 w_x+ 6 u_x^2+ 6 u u_{xx} + 6 v u_x  + 
6 u v_x  - 12 u^2 u_x = 0.
\end{eqnarray}
Eq.(\ref{kp}) 
 results in the integrated KP after elimination of $v$ and $w$:
\begin{eqnarray}
u_{t_3} - \frac{1}{4} u_{xxx} -
 \frac{3}{4} \partial^{-1} u_{t_1t_1} + 3 u_x^2 = 0
\end{eqnarray}
For this reason we call the above system (\ref{u21}) - (\ref{u26}) 
(together with the whole list of nonlinear PDE 
related with different variables $t_\gamma$) the generalization of  KP
hierarchy. 

\section{Construction of particular solutions}

For  construction of particular solutions to the nonlinear PDE 
which appear in this paper, one needs to choose the functions 
$\psi_i$, $i=1,\dots, Q_{n-1}$  which satisfy the conditions 
(\ref{det}) and (\ref{t}).
One of the following Fourier integrals is suitable for this
purpose.

For the equations from the Sec.2:
\begin{eqnarray}\label{F1}
\psi_i=\int c_{i}({\bf k}) \exp\left(\sum_{i=1}^N k_i x_i + 
  \sum_{\gamma} \omega_\gamma({\bf k}) t_\gamma \right) d {\bf k},\\
\nonumber
{\bf k}=(k_1,\dots,k_N),\;\;i=1,\dots,Q_{n-1}.
\end{eqnarray} 
Here  
$\omega_\gamma$ depends on ${\bf k}$ due to the dispersion equations associated with the eqs.(\ref{t})
\begin{eqnarray}
\omega_\gamma = k^\gamma.
\end{eqnarray}
In this case $\psi_i$ are arbitrary functions of all variables $x_i$. 

For the equations from the Sec.3:
\begin{eqnarray}\label{F2}
\psi_i=\int c_{i}(k_1) \exp\left(k_1 x_1 + \sum_{i=2}^N k_i(k_1) x_i + 
  \sum_{\gamma} \omega_\gamma( k_1) t_\gamma \right) d 
  k_1,\\\nonumber
  \;\;i=1,\dots,Q_{n-1}.
\end{eqnarray} 
Functions $\omega_\gamma (k_1)$ and $k_i(k_1),\;\;i>1$ 
represent the dispersion relations associated with eqs.(\ref{t})
and (\ref{x}):
\begin{eqnarray}
\omega_\gamma = k^\gamma,\;\;k_i^{n_i+1} = 
\prod_{j=1}^N k_j^{\alpha^{(i)}_j},\;\; a^{(i)}_i=0,\;\;
i=2,3,\dots,N.
\end{eqnarray}
Here $\psi_i$ are arbitrary functions of single variable $x_1$;
$c_{i}({\bf k})$ and $c_i(k_1)$ in formulas (\ref{F1}) and (\ref{F2}) 
are such functions of 
argument(s) that condition (\ref{det}) is satisfied.
 
After this, one needs to solve the system of algebraic equations 
(\ref{Su1}) for coefficients $w^\alpha_\beta$ 
 which are the solutions of appropriate system of nonlinear PDE  (\ref{Su2}). 

For instance, let us construct the particular solution for the system 
(\ref{uvw}). In this case the equation (\ref{F2}) should 
 be written in the form
\begin{eqnarray}
\psi_i=\int c_{i}(k_2) \exp\left(k_2^2 x_1 + k_2 x_2 + 
k_2^4 t_1 \right) dk_2,\;\;i=1,\dots,Q_{n-1}.
\end{eqnarray}
To construct the simple solution, let $Q_{n-1}=3$ (or $n=2$ due to the
eq.(\ref{Q})) and  
 take the following expressions for $\psi_i$:
\begin{eqnarray}
\psi_i=c_{1i} + c_{2i} \exp(k_i^2 x_1 + k_i x_2 + k_i^4 t_1),\;\;i=1,2,3. 
\end{eqnarray}
Then the systems (\ref{Seq2i}) and (\ref{Seq2ii}) 
becomes $3\times 3$ algebraic matrix equations 
for the coefficients $u_{ij}$ and $v_{ij}$ respectively, 
which can be solved directly. One can see that the condition 
(\ref{det2}) is satisfied for our choice of functions $\psi_i$. 
We give only expression for  the function 
 $u=v_{00}$:
\begin{eqnarray}
u&=&\frac{s_1 E_1 E_2 +s_2 E_1 E_3 +s_3 E_2 E_3+s_4 E_1 E_2 E_3 }
{p_1 E_1 E_2 +p_2 E_1 E_3 +p_3 E_2 E_3+p_4 E_1 E_2 E_3},\\\nonumber
E_i&=&\exp(k_i^2 x + k_i y + k_i^3 t_1),\;\;\\\nonumber
s_1&=& c_{13} c_{21} c_{22} k_1 k_2 (k_2^2-k_1^2),\;\;
s_2\;=\; c_{12} c_{21} c_{23} k_1 k_3 (k_1^2-k_3^2),\;\;\\\nonumber
s_3&=& c_{11} c_{22} c_{23} k_2 k_3 (k_3^2-k_2^2),\;\;\\\nonumber
s_4&=& c_{21} c_{22} c_{23} k_1 k_2 (k_2-k_1)(k_1-k_3)
   (k_2-k_3)(k_1+k_2+k_3),\;\;\\\nonumber
p_1&=&c_{13} c_{21} c_{22} k_1 k_2 (k_1-k_2),\;\;
p_2\;=\;c_{12} c_{21} c_{23} k_1 k_3 (k_3-k_1),\;\;\\\nonumber
p_3&=&c_{11} c_{22} c_{23} k_2 k_3 (k_2-k_3),\;\;
p_4\;=\;c_{21} c_{22} c_{23} (k_1-k_2)(k_1-k_3) (k_2-k_3),\;\;
\end{eqnarray}
If all coefficients $p_i$ in the denominator of this solution 
have the same sign (this can be arranged by the appropriate choice
of parameters $c_i$ and $k_i$), 
then the solution has no singularities for all 
values of independent variables and describes inelastic three-kink 
interaction.

\section{Conclusions} 

We represent an example of
 multidimensional generalization of Sato equation (\ref{SB}), 
written in terms of pure differential operators. 
We give algorithm for construction of 
different  multidimensional  hierarchies  which admit the
special family of particular solutions. Namely, solutions are parameterized by the set of arbitrary functions of  either one or several  variables $x_i$.
  Different subspaces of general space 
 (\ref{W}) result in specific systems of nonlinear PDE. Two examples
 of this kind are given in the  Sec.3.

The represented approach might be considered 
as  the mixture of Sato theory and Lax 
representation. In fact, each equation of the system (\ref{SB}) is 
nothing but the compatibility 
condition of the overdetermined linear systems (\ref{t}) and (\ref{Seq}).  
For each particular $\gamma$,
the equation (\ref{SB}) 
 gives the complete system of nonlinear PDE (\ref{Su2}) on 
the coefficients $w^\alpha_\beta$ of the operators $W_\alpha$.
But the structure of this 
 system 
depends on $n$ (order of differential operators $S_\alpha$) in general case. 
The examples of systems of nonlinear PDE which do not depend on $n$ 
are given in the Sec.3. 

Author thanks Professor A.H.Zimerman (IFT UNESP, Sao P\~aulo, Brazil) for   
initiation of this work. The work 
is supported by RFBR grants 01-01-00929 and 00-15-96007.


\begin{thebibliography}{30}

\bibitem{GGKM}
C.S.Gardner, J.M.Greene, M.D.Kruskal and R.M.Miura, Phys.Rev.Lett.,
{\bf 19}, 1095 (1967)

\bibitem{AKNS}
M.J.Ablowitz and H.Segur, {\it Solitons and Inverse Scattering Transform}, 
(SIAM, Philadelphia, 1981)


\bibitem{ZMNP}
V.E.Zakharov, S.V.Manakov, S.P.Novikov and L.P.Pitaevsky, 
{\it Theory of Solitons. The Inverse Problem Method}, 
(Plenum Press, 1984)

\bibitem{WH}
G.B. Whitham, {\it Linear and Nonlinear Waves} (Wiley, NY, 1974).

\bibitem{CH1}
 R.Camassa and D.D.Holm, Phys.Rev.Lett. {\bf 71}, 1661 (1993)

\bibitem{CH2}
 M.S.Alber, R.Camassa, D.D.Holm and J.E.Marsden, 
Lett.Math.Phys. 
 {\bf 32}, 137 (1994) 


\bibitem{FO}
A.S.Fokas, Physica D {\bf 87}, 145 (1995)
 
\bibitem{FF}
B.Fuchssteiner and A.S.Fokas, Physica D {\bf 4}, 47 (1981) 

\bibitem{CGP}
P.A.Clarkson, P.R.Gordoa and A.Pickering, Inv. Prob. {\bf 13}, 1463 (1997)

\bibitem{HKO}
A. Hasegawa and Y. Kodama, 
{\it Solitons in Optical Communication}, Oxford Univ. Press (1995)

\bibitem{A}
G.P.Agrawal, {\it Nonlinear Fiber Optics}, Acad. Press (1994)

\bibitem{OSTT}
 Y.Ohta, J.Satsuma, D.Takahashi and T. Tokihiro,
{\it An elementary introduction to Sato Theory}, Progr. Theor.Phys. Suppl.,
No.94, p.210 (1988).

\bibitem{M} F.J.Plaza Mart\'in, math.AG/0008004

\bibitem{P}A.N.Parshin, Proc. Steklov Math.Inst., {\bf 224}, 266 (1999)

\bibitem{H1}
R.Hirota, in 
Lecture Notes in Mathematics {\bf 515}, Springer-Verlag, New York (1976)

\bibitem{H2}
R.Hirota, J.Phys.Soc.Japan, {\bf 46}, 312 (1979)


\bibitem{HS}
R.Hirota and J.Satsuma, Progr.Theoret.Phys.Suppl., {\bf 59}, 64 (1976)

\bibitem{HS2}
R.Hirota and J.Satsuma, J.Phys.Soc.Japan, {\bf 40}, 891 (1976)

\bibitem{WTC}
J.Weiss, M.Tabor and G.Carnevale, J.Math.Phys., {\bf 24}, 522 (1983)

\bibitem{W}
J.Weiss, J.Math.Phys., {\bf 24}, 522 (1405)

\bibitem{EG}
P.Estevez and P.Gordoa, J.Phys.A:Math.Gen., {\bf 23}, 4831 (1990) 


\bibitem{Z1}
A.I.Zenchuk, Phys.Lett.A, {\bf 277}, 
(2000) 25

\bibitem{Z2}
 A.I. Zenchuk, J.Math.Phys.,{\bf 42}, 5472 (2001)

\bibitem{Z3}
A.I.Zenchuk,  J.Phys.A:Math.Gen. {\bf 35}, 1791 (2002) 

\end{thebibliography}
\end{document}